\documentclass[12pt]{article}
\newcommand{\beq}{\begin{equation}}
\newcommand{\eeq}{\end{equation}}
\newcommand{\beqa}{\begin{eqnarray}}
\newcommand{\eeqa}{\end{eqnarray}}
\newcommand{\half}{\frac{1}{2}}
\newcommand{\Gsim}{\mbox{\raisebox{-.3em}{$\stackrel{>}{\sim}$}}}
\newcommand{\Lsim}{\mbox{\raisebox{-.3em}{$\stackrel{<}{\sim}$}}}
\newcommand{\Lmd}{\Lambda}
\newcommand{\MP}{M_{\rm P}}

\newcommand{\reflef}{(\ref}

\begin{document}
\begin{center}
{\Large\bf Is the zero-point energy a source of the cosmological constant? }
\mbox{}\\[1.0em]

{\large\bf Yasunori Fujii}\\

Advanced Research Institute for Science and Engineering, Waseda University, Okubo, Shinjuku-ku, Tokyo, 169-8555, Japan
\mbox{}\\[.5em]
Email fujiitana@gmail.com
\mbox{}\\[.5em]
\begin{abstract}

We discuss how we remove a huge discrepancy between the theory of a cosmological constant, due to the zero-point energies of matter fields, and the observation. The technique of dimensional regularization plays a decisive role.  We eventually reach the desired behavior of the vacuum  densities falling off $\sim t^{-2}$, allowing us to understand how an extremely small result comes about naturally.  As a price, however, the zero-point energy vacuum fails to act as a true cosmological constant.  Its expected role responsible for the observed accelerating universe is then to be inherited by the gravitational scalar field, dark energy, as we suggest in the scalar-tensor theory.

\end{abstract}
\end{center}
\mbox{}\\[-3.6em]

\section{Introduction}

According to a traditional idea of the relativistic quantum field theory, each  field is associated with the zero-point energy vacuum to be identified eventually with the cosmological constant, CC \cite{wb,cpt}.  The calculated CC in this way turns out, however, to be extremely larger, by as much as 50 orders of magnitude or even larger,  than the observed value of CC determined to fit the way of the acceleration of the universe \cite{accelu}.  From a theoretical point of view, on the other hand, the issue is also unique in the sense that such fundamental concepts like Lorentz invariance combined with dimensional regularization DR are challenged directly  by a simple argument. This might allow the scenario to be developed that  we start with assuming a negligibly small  seed  of Lorentz non-invariance, mounted on tangentially flat spacetime, to grow into a cosmological effect of barely measurable size, taking place in curved spacetime for the expanding universe.

In our present short article, however, we are going to limit ourselves to a {\em toy model}, in which no other matter density is included in addition to the zero-point energy state, basically, for the sake of comparison with the way in which  the issue had been discussed originally in  the related portions of \cite{wb,cpt}.

In Sec.\hspace{.3em}2, we explain the minimized basics of the technique of DR as required later, first in 4 dimensional flat Minkowski spacetime ${\cal M}_0$.  We show then how we move to curved spacetime $\cal C$
 for the cosmology of an expanding universe.  We come to finding how desperately the simplest theoretical result on the vacuum energy density $\rho_v ={\rm const}$ differs from the observed CC.

As a remedy, in Sec.\hspace{.3em}3, we start with introducing ${\cal M}$, a non-Minkowskian, or Lorentz non-invariant, extension of ${\cal M}_0$, deriving {\em different} cosmological solutions,  most remarkably featuring the behavior $\rho_v\sim t^{-2}$, naturally reaching the desired small $\rho_v$.

After more detailed analyses including the size of $\rho_v$ we  find it quite likely that the effect is {\em hidden} deep below the observed CC, in agreement thus with the observation.  This is what we are going to mean by a {\em success} of the proposed theoretical approach in evading the immediate difficulty as claimed by \cite{wb,cpt}.

The final Sec.\hspace{.3em}4 is devoted to discussing other consequences.  A particular emphasis is placed on the importance of the gravitational scalar field, the dark energy, in the scalar-tensor theory STT, which provides us with a better candidate of an effective CC to implement an observed accelerating universe.

\section{The simplest theory confronted with the observation}
\setcounter{equation}{0}

For simplicity, we confine ourselves to a free massive  neutral scalar field $\Phi$, as a representative of the matter, in Minkowski spacetime ${\cal M}_0$ with the Lagrangian\footnote{Our estimate throughout the current article is based on the (reduced) Planckian units: $c=\hbar=\MP(=(8\pi G)^{-1/2})=1$.  The unit of length, time, energy in this unit-system expressed in the conventional units are $8.10 \times 10^{-33}{\rm cm}, 2.70 \times 10^{-43}{\rm sec}, 2.44 \times 10^{18}{\rm GeV}$, respectively. The present age of the universe is $t_0\approx 1.37 \times 10^{10}{\rm y}\approx 1.59 \times 10^{60}=10^{60.2}$. }
\beq
L =-\half \eta^{\mu\nu}\partial_\mu\Phi \partial_\nu\Phi-\half m^2 \Phi^2,
\label{zptear3_5}
\eeq
with the energy-momentum tensor;
\beq
T_{\mu\nu}=\partial_\mu\Phi \partial_\nu\Phi+\eta_{\mu\nu}L.
\label{zptear3_6}
\eeq

We compute the vacuum expectation values for the energy- and pressure-densities;
\beqa
\rho_v=<\rho>&=& <T_{00}>=(2\pi)^{-3}\half\int d^3k\cdot \omega(k), \label{reg_1}\\
p_v=<p>&=& \frac{1}{3}\delta^{ij}<T_{ij}>=(2\pi)^{-3}\half\frac{1}{3}\int d^3k \frac{k^2}{\omega(k)},
\label{reg_1a}
\eeqa
where $\omega(k)=\sqrt{k^2+m^2}$.

Notice that we integrate over the phase volumes represented by the three-momenta $k$.  We then face, as was emphasized by Martin in IV of \cite{mart}, and other references cited therein, an immediate difficulty that the integrals depend too much on the cutoff momenta $K$; $\rho_v\approx K^4/(16\pi^2), p_v\approx K^4/(48\pi^2)$ for $K\gg m$, also far from giving the ratio $p_v/\rho_v=-1$ expected for the vacuum.  He continues further to indicate that the only remedy comes from  applying the  DR technique to the integral  together with the factor $1/3$ in \reflef{reg_1a}) replaced by $1/(D-1)$, 
\beqa
\rho_v &=&(2\pi)^{-D+1}\half\int d^{D-1}k\cdot\omega(k), 
\label{reg_2} \\
p_v&=&(2\pi)^{-D+1}\half \frac{1}{D-1}\int d^{D-1}k\frac{k^2}{\omega(k)},
\label{reg_3}
\eeqa
where the physical value is $D=4$.

After certain straightforward  details \cite{yfkm},\footnote{Appendix N, in particular.  To simplify the core of the argument, we will  omit the multiplicative factors of the type $\ln (m^2/\mu^2)$.  See \cite{mart,cher}, for more details.} we obtain
\beqa
\rho_v&=&m^D(2\pi)^{-D+1}V_{D-1}\frac{\Gamma(d-\half)\Gamma(-d)}{\Gamma(-\half)},
\label{zptear3_7}\\
p_v&=&m^D(2\pi)^{-D+1}V_{D-1}  \frac{\Gamma(d+\half)\Gamma(-d)}{\Gamma(\half)}\frac{1}{D-1},
\label{zptear3_8}
\eeqa
where
\beq
V_{D-1}=\frac{2\pi^{d-1/2}}{\Gamma (d-1/2)},\quad\mbox{with}\quad  d=\frac{D}{2}.
\label{zptear3_9}
\eeq
Both of $\rho_v$ and $p_v$ share the pole $\Gamma(-d)=(-d)^{-1}\Gamma(1-d)=(-d)^{-1}(1-d)^{-1}\Gamma(2-d)\approx \frac{1}{2}(2-d)^{-1}$.  Nevertheless the ratio turns out to be finite;\footnote{Sometimes another symbol $w$ is used for the same ratio $z$.}
\beq
z=\frac{p_v}{\rho_v}=-1,
\label{zptear3_10}
\eeq
for {\em any} values of $D$, a {\em global} nature in 1-dimensional space of $D$, also thanks to the relations,
\beq
\hspace{-3.0em}\Gamma\left(d+\half\right)=\left( d-\half \right)\Gamma\left(d-\half\right),
\quad\mbox{and}\quad
\sqrt{\pi}=\Gamma\left(\half\right)=-\half\Gamma\left(-\half\right).
\label{zptear3_11}
\eeq
As we point out, the minus sign in \reflef{zptear3_10}), inheriting originally the Lorentz invariant result $<T_{\mu\nu}> \sim\eta_{\mu\nu}$, comes from the same in the second of \reflef{zptear3_11}), representing explicitly an admitted violation of naive positivity of the integrand as was discussed in \cite{yfmm}, one of the unique features   in the computation of the gamma functions which are the core ingredients of the technique of DR.  In this respect, the use of DR is of special importance, like gauge invariance emphasized in other areas, like QCD, for example \cite{muta}.

 From \reflef{zptear3_7}) and \reflef{zptear3_8}), we naturally find that $\rho_v$ and $p_v$ stay constant separately;
\beq
\dot{\rho}_v =0, \quad \dot{p}_v =0.
\label{zptear3_12}
\eeq

We might then move  to curved spacetime ${\cal C}$ for the expanding universe.  The conservation law \reflef{zptear3_12}) will be replaced by
\beq
\dot{\rho}_v +(D-1)H{\cal F}\rho_v=0,\quad\mbox{where}\quad {\cal F}=1+z,
\label{zptear3_13}
\eeq
with $H=\dot{a}/a$, also assuming a spatially flat $(D-1)$-space for the Robertson-Walker metric.\footnote{The  covariant  conservation laws, or the Bianchi identities are; $\nabla_\nu T^{\nu}_\mu =(1/\sqrt{-g})\partial_\nu(\sqrt{-g}T^\nu_\mu)-\!(1/2)(\partial_\mu g_{\rho\sigma})T^{\rho\sigma}\!\!=0$, with $T_{00}=\rho, T_{ij}= pg_{ij}$.  The required energy conservation law is for $\mu=0$. }   Using \reflef{zptear3_10}) makes the second term of \reflef{zptear3_13}) to vanish for whatever $H$ and $D$, hence deriving
\beq
\dot{\rho}_v =0,
\label{zptear3_14}
\eeq
the same as the first of \reflef{zptear3_12}), leaving us in conflict with the observation, to be outlined briefly.

According to \reflef{zptear3_7}) we have $m^4$
 though it is multiplied by an infinity coming from $\Gamma(2-d)$.  Suppose this pole term can be subtracted away to
 be replaced by a  finite number \cite{muta}.
Then we are going to have
\beq
\rho_v \sim \Lmd_{\rm eff} \sim m^D\sim m^4.
\label{zptear3_15}
\eeq
Notice that, uniquely with DR, this $m$ is,
in accordance with \reflef{zptear3_5}),   the mass of the field, 
which makes up the vacuum, instead of the cutoff  $K$ as mentioned before.

As a tentative choice, suppose $m$ is typically $\sim {\rm GeV}\sim 10^{-18}$ in the Planckian units to
 give $\Lmd_{\rm eff} \sim 10^{-72}$ which is still too large compared with the observed value \cite{accelu}
;\footnote{Strictly speaking, $\Lmd_{\rm obs}$ on LHS implies the component corresponding to the
 truly constant CC, or dark energy, which occupies about $3/4$ of the total energy density.    Any component,
 including the purely non-CC part, remains roughly of the same order of magnitude.  Also to be emphasized, 
our  finding $10^{-120}$ to be close unmistakably to $t_0^{-2}$, hardly coincidentally, has been a 
major support to  the scalar-tensor theory STT \cite{yfkm,Entr,yfkh}. }
\beq
\Lmd_{\rm obs} \sim 10^{-120} \sim t_0^{-2},
\label{zptear3_15a}
\eeq
by as much as nearly 50 orders
 of magnitude.  The two equations could have been consistent with each other, only if $m$ were to be chosen as small as $\sim 10^{-120/4}\sim 10^{-30}\sim 10^{-12}{\rm GeV}\sim 10^{-3}{\rm eV}
={\rm meV}$, unrealistically light for ordinary elementary particles; any  particle heavier than this would contribute too much to CC.
This issue, in spite of the past efforts,\footnote{For example, the lack of complete mass-degenerations between bosons and fermions, allows us no super-symmetric cancellations in \reflef{zptear3_15}) to occur up to the required level of order-of-magnitude estimate. } is still one of the 
most serious puzzles on CC, but might be evaded only if $\rho_v$ does {\em decay} with time, certainly a necessary  condition. This  forces us to choose ${\cal F} \neq 0$ on LHS of \reflef{zptear3_13}). We are then led to go through another non-trivial step, as will be shown.

\section{Partially regularized densities and cosmology}
\setcounter{equation}{0}

In Sec.\hspace{.3em}2 we suggested to find ourselves living in 1-dimensional space of $D$, with a special point $D=4$ corresponding to the physical world.  After a straightforward computation in ${\cal M}_0$, we have reached \reflef{zptear3_10}) together with the brief  comment following immediately, finding a uniformly flat distribution ${\cal F}=0$ everywhere, in sharp contradiction with its non-vanishing in the necessary condition stated above. 
To close an obvious gap, we believe it worth attempting to start with assuming Lorentz non-invariance in ${\cal M}$, flat spacetime off $D= 4$ as an extension of ${\cal M}_0$, with the corresponding terms kept small enough to be consistent with probably none or little of the  evidences reported so far.  We then move to curved spacetime ${\cal C}$ making use of \reflef{zptear3_13}), expecting cosmological effects, which turn out sufficiently sizable in a theoretically natural manner.

As a general idea, we have a relatively firm  confidence for $D=4$, including the Lorentz invariance hence ${\cal F}=0$, while we have much less certainty for $D\neq 4$.\footnote{Technique of DR provides us with a strong support for the presence of extra dimensions, but only theoretically, lacking more unequivocal experimental tests.}  In accordance with this aspect, we might be led naturally
to assuming $\cal F$ to be  a function of $D$ such that  
\beq
{\cal F} \rightarrow 0, \quad \mbox{as}\quad D\rightarrow 4.
\label{zptear3_15b}
\eeq
It is even more interesting to find this simplified but heuristic description  to be implemented by a simple function
\beq
{\cal F}=1-f_p,
\label{zptear3_15c}
\eeq
where
\beq
\hspace{-3.0em}f_p(D)\equiv \frac{D-1}{3}=\frac{2}{3}\left( d-\half \right) =1+\frac{2}{3}\epsilon,\quad\mbox{where}\quad\epsilon =d-2=\half (D-4),
\label{ftzpte_2}
\eeq
allowing us to re-express \reflef{zptear3_15b}) as
\beq
{\cal F}= -\frac{2}{3}\epsilon,\quad\mbox{as}\quad \epsilon \rightarrow 0, \quad\mbox{or}\quad D\rightarrow 4.
\label{zptear3_15d}
\eeq
Basically, we are preparing a {\em seed} of a Lorentz non-invariance in $\cal M$, which is away from the original Minkowski ${\cal M}_0$ in 4 dimensions, only by the {\em negligibly} small amount $|\epsilon |\ll 1$.

Before entering into details on the consequences of \reflef{ftzpte_2}) and \reflef{zptear3_15d}), we are going to show that \reflef{ftzpte_2}) has an intuitive  implication bridging between \reflef{reg_3}) and \reflef{reg_1a}).  The former is derived from the fully $D$-dimensionalized Lagrangian, while the latter, the {\em partially regularized density},  agrees with the former only for $D=4$, though with the loop integral still regularized for any $D$.   Also to be recalled, \reflef{reg_3}) corresponds to using the spatial metric components $\eta_{ij} =\delta_{ij}$, while the result with \reflef{zptear3_15c}) and \reflef{ftzpte_2}) is re-interpreted by replacing it by $g_{ij}=\delta_{ij}f_p$, a different geometry.  Obviously, \reflef{ftzpte_2}) is only a simple example, to be modified for more complications, which might be found in the future.  In the current starting attempt, however, we focus almost exclusively  on the illustrative choice \reflef{ftzpte_2}), to avoid unnecessary complications.

We now substitute \reflef{zptear3_15d}) back to \reflef{zptear3_13}) with $D-1$  replaced safely by the factor 3.  The result  is in fact $t$-independent;  
\beq
\frac{d\rho_v}{\rho_v}=-3{\cal F}\frac{da}{a},
\label{eq_2}
\eeq
to be  integrated immediately giving
\beq
\rho_v=\rho_{v0}a^{-3{\cal F}},
\label{eq_3}
\eeq
with $\rho_{v0}$ an integration constant.  Substituting this into the Einstein equation
\beq
\frac{\dot{a}}{a}=\frac{1}{\sqrt{3}}\rho_v^{1/2},
\label{eq_1}
\eeq 
giving
\beq
\frac{\dot{a}}{a}=\frac{1}{\sqrt{3}}\rho_{v0}^{1/2}a^{-(3/2){\cal F}},
\label{eq_4} 
\eeq
which is again integrated with respect to $t$ to yield
\beq
t-t_{\rm c}=\sqrt{3}\rho_{v0}^{-1/2}\frac{2}{3}{\cal F}^{-1}a^{(3/2){\cal F}},\quad\mbox{or}\quad a^{(3/2){\cal F}}=\frac{1}{\sqrt{3}}\rho_{v0}^{1/2}\frac{3}{2}{\cal F}\cdot(t-t_{\rm c}),
\label{eq_5}
\eeq
where $t_{\rm c}$ is another integration constant.
Eliminating $a^{(3/2){\cal F}}$ from \reflef{eq_3}) and the second of \reflef{eq_5}), we derive
\beq
\rho_v=\rho_{v0}\left( a^{(3/2){\cal F}} \right)^{-2}=\frac{4}{3}{\cal F}^{-2}t^{-2}=3 \epsilon^{-2}t^{-2},
\label{zptear3_19y}
\eeq
where use has been made of \reflef{zptear3_15d}), with $t_{\rm c}$ dropped. Notice also that the constant $\rho_{v0}$ in the second equation in \reflef{zptear3_19y}) has been cancelled by the same which occurs in \reflef{eq_3}).
 Now the solution \reflef{zptear3_19y}) shows two remarkable features; the time-dependence $\sim t^{-2}$ and the {\em singular} behavior $\epsilon^{-2}$ as $\epsilon \rightarrow 0$.  Their implications will be dissussed separately.

\subsection{Time-dependence}

The time-dependence $\sim t^{-2}$ which occurs in \reflef{zptear3_19y})  is unique and universal independently of the parameter $\epsilon$,\footnote{This feature comes directly from the explicit absence of $t$ in $\cal F$, as in conventional equations of state, like the radiation- and dust-dominated universes.  } indicating a consistency with the observation \reflef{zptear3_15a}).

One of the convenient ways to probe the time-variability of $\rho_v(t)$ is to {\em compare} the density at two different times, $t_1$ and $t_2$,  as represented by the {\em ratio} 
\beq
{\cal R}(t_1, t_2)=\frac{\rho_v(t_2)}{\rho_v(t_1)}=\frac{\hat{\rho}_v(t_2)}{\hat{\rho}_v(t_1)}=\left( \frac{t_1}{t_2} \right)^2 ,
\label{zptear7_5}
\eeq
which is  independent of the multiplicative factor $\epsilon^{-2}$ present in the density as given by \reflef{zptear3_19y}), with the symbol $\:\hat{}\:$ specifically for its dropping.  As a special caution, however, we always  start with  keeping $\epsilon$ nonzero finite, in order to define the ratio unambiguously, also following the rule of DR.

As a relevant example, we may choose $t_1=1$ for a unit time in the very early universe at which we have two typically different  {\em theoretical} estimates;
\beq
\hat{\rho}_v(t_1)=\Lmd_1 \sim \left\{
\begin{array}{l}
1,  \\[.8em]
m^4 \sim 10^{-72},　\quad\mbox{for}\quad m\sim {\rm GeV},
\end{array}
\right.
\label{zptear7_5a1}
\eeq                                    
with the first line for a {\em naive} choice, while the second one  corresponding to a tentative  choice particularly suited to  DR, as discussed following  \reflef{zptear3_15}).  
According to \reflef{zptear7_5}) we then find today's values;
\beq
\hat{\rho}_v(t_0)= \Lmd_0 =\Lmd_1 \left( \frac{t_0}{t_1} \right)^{-2}\sim\left\{
\begin{array}{l}
t_0^{-2},  \\[.8em]
m^4 t_0^{-2}\sim 10^{-72}t_0^{-2},
\end{array}
\right.
\label{zptear7_5a2}
\eeq
confirming the agreement with \reflef{zptear3_15a}) with $\Lmd_{\rm obs}\sim \hat{\rho}_v(t_0)$ as far as  the way of time-dependence is concerned.   The multiplicative factor will be discussed below.

\subsection{Size of the density }

Due to the presence of $\epsilon^{-2}$, we have the solution \reflef{zptear3_19y}) only for $\epsilon \neq 0$, which originally parametrizes the extent of Lorentz non-invariance prepared in ${\cal M}$.  In this flat spacetime, in contrast, 
 we have only smooth behaviors around $\epsilon =0$ including the point corresponding to $D=4$.  We might find out other singularity-free theories even for curved spacetime.  At this moment, however, we might fortunately take advantage of understanding that our universe had chosen a nonzero $\epsilon$.

We then use \reflef{zptear7_5a2}) multiplied by $\epsilon^{-2}$,  to derive $\rho_v(t_0)$ today;
\beq
\rho_v(t_0)
\sim\left\{
\begin{array}{l}
\epsilon^{\: -2}t_0^{-2},  \\[.8em]
\epsilon^{\: -2}m^4 t_0^{-2},
\end{array}
\right.
\label{zptear7_5a22}
\eeq
which is further substituted into \reflef{zptear3_15a}) with $\rho_v \Lsim \Lmd_{\rm obx}$ on LHS, resulting in the inequalities; 
\beq
\epsilon^{\:-2}\;\Lsim \;\left\{
\begin{array}{l}
1,  \\[.8em]
m^{-4}, 
\end{array}
\right.
\quad \mbox{or}\quad |\epsilon |\;\Gsim \;\left\{
\begin{array}{l}
1, \\[.8em]
m^2 \sim 10^{-36}.
\end{array}
\right.
\label{zptear7_c3}
\eeq

The first line is unacceptable because of the required smallness of $|\epsilon | <1$, while the second line might signify a {\em success} of the approach because it stays smaller than what is supposed to be today's amount of dark energy, thus 
keeping the ensuing phenomena to be consistent with the observations.  The above same estimate also allows a surprisingly small lower bound of $|\epsilon|$, illustrating how a negligibly small seed, mentioned before, may grow into a cosmological effect of barely measurable size.   As an extreme example, with $|\epsilon |$ as {\em small} as $\sim m^2 \sim10^{-36}$, to be enhanced by $\epsilon^{\:-2}$ in \reflef{zptear3_19y}),   we might reach today's upper bound as {\em large} as $\rho_v \sim t_0^{-2}$.

In more detail, suppose there is the heaviest fundamental particle, the Higgs field of the mass $\sim 126\hspace{.2em}{\rm GeV}$ \cite{higgs}, for example, hence $m_{\rm H}^2\sim 10^{-32}$, which acts as a lower bound of $|\epsilon|$ commonly to any of other particle masses of smaller $m$'s.

We may also point out that we have reached  the {\em success} only within our {\em toy model}, basically the same as in \cite{wb,cpt}.  We should still wait until the same conclusion emerges essentially unchanged when the same type of analysis is applied to more realistic situations with other matter densities included.

Also as we mentioned in \reflef{ftzpte_2}), our foregoing analyses in terms of $\epsilon$ can be re-expressed directly in terms of 
\beq
D =4+2 \epsilon,
\label{ex_24}
\eeq
to be called {\em non-integer dimensionality} NID, rather than many other theoretical proposals based so far on integer-dimensional spacetime.

We are then urged to inquire how NID can be justified numerically.  As a simple yet overall way, we focus upon the massless static gravitational potential given by $r^{-b}$   where $b=D-3=1+2\epsilon$, with its relative change approximated by $\Delta =(r^{-1})^{-1}(r^{-(1+2\epsilon)}-r^{-1}) \sim-2\epsilon\ln r \sim 2\epsilon$.  This might be compared with the result of the precision measurements to determine to what extent Newtonian gravity, starting with the issue of the inverse-square law \cite{Nat} expected from the exchange of the dilaton-like scalar field, can be verified  to the accuracy of $\sim 10^{-10}$\cite{nNewt}.

 Our constraint $\epsilon \sim 10^{-36}$ mentioned above in \reflef{zptear7_c3})
 is far more stringent than \cite{nNewt}, at least nominally, though a real comparison between the two attempts of vastly different nature might be highly non-trivial.  As we find, on the other hand, even with a {\em reasonably} small value $\epsilon \sim 10^{-10}$, as suggested crudely by \cite{nNewt},  we have arrived, in accordance with the second lines of \reflef{zptear7_5a2}) and \reflef{zptear7_c3}), at $\rho_v \sim 10^{20-72}t_0^{-2}\sim 10^{-52}t_0^{-2} \ll t_0^{-2}$, sinking quite deeply invisible.

\section{Other consequences}
\setcounter{equation}{0}

We  put a particular emphasis on the way of time-dependence $\sim t^{-2}$,  because this would represent precisely the way in which the huge discrepancy mentioned before between the theory and the observation is going to be removed in a manner fully  consistent with the Scenario of a decaying CC, as developed in our scalar-tensor theory STT \cite{yfkm,Entr,yfkh}.  In the latter approach, the smallness of the observed CC is understood naturally in terms of an old age of the universe, hence leaving us free from the fine-tuning problem, as shown by \reflef{zptear3_15a}).

A simple falling-off behavior as shown above, however, comes with a price; it fails to provide us with a truly constant CC, which is supposed to cause a mini-inflation of the ordinary matter, or the accelerating universe as we see it \cite{accelu}.  This effect, as we suggest, might be entirely due to our STT, in which we have  the gravitational scalar field, not to be confused with the scalar field $\Phi$ as discussed  in \reflef{zptear3_5}) and thereafter as a simplified representative of matter fields.  We exploited a nonlinear nature of the cosmological equations in which the gravitational scalar field density, dark energy, shows a {\em sporadic} leveling-off behavior effectively playing the role of a truly constant CC for {\em limited} time durations, supported also by what is called a trapping mechanism \cite{yfkm,Entr}.  We also expect to include the contribution from the zero-point energy state to improve the status of the {\em toy model} in the current analysis.

As another aspect, we no longer expect the semi-classical interpretation of some of the radiative corrections \cite{welt}, due to the vacuum fluctuation of the zero-point energies of the photons, applied to the Lamb-shift \cite{mart,kroll}, or the anomalous magnetic moment of the electron \cite{koba}, for example.  These results should be re-examined in view of $\rho_v=0 $  for the massless photons, derived basically from \reflef{zptear3_7}) as a unique result of DR \cite{mart,cher}.   We might even raise a somewhat cautious question if the semi-classical approach plays a better role than the fully relativistic and quantized field-theoretical calculations.
\\[.5em]

 \noindent
{\Large\bf Acknowledgements}
\mbox{}\\[.5em]

The author wishes to express his sincere thanks to Yuichi Chikasige, Mamoru Doi,  Masakatsu Fujimoto, Kensuke Homma, Akio Hosoya, Akira Iwamoto, Susumu Kamefuchi, Kazuaki Kuroda, ChongSa Lim, Kei-ichi Maeda, Tsukasa Tada, Jun'ichi Yokoyama, Tamiaki Yoneya, and Satoshi Watamura for their helpful discussions.

\small{
 
}

\end{document}